\documentclass[%
 reprint,
 amsmath,amssymb,
 aps,
]{revtex4-1}
\usepackage{xcolor}
\usepackage[normalem]{ulem}
\usepackage{graphicx, color, braket}% Include figure files
\usepackage{dcolumn}% Align table columns on decimal point
\usepackage{bm}% bold math
\usepackage[colorlinks=true,linkcolor=blue]{hyperref}
\renewcommand*{\Im}{\operatorname{Im}}
\renewcommand*{\Re}{\operatorname{Re}}
\newcommand{\New}[1]{\textcolor{black}{#1}}
\newcommand{\new}[1]{\textcolor{black}{#1}}
\renewcommand*{\eta}{\nu}

\begin{document}

\preprint{APS/123-QED}

\title{Ballistic-to-hydrodynamic transition and collective modes \\
for two-dimensional electron systems in magnetic field}

\author{Kirill Kapralov$^{1,2}$}
 \email{kapralov.kn@phystech.edu}
\author{Dmitry Svintsov$^1$}%
\affiliation{%
 $^1$Center for Photonics and 2d Materials, Moscow Institute of Physics and Technology (National Research University), Dolgoprudny 141700, Russia. \\
 $^2$Kotelnikov Institute of Radioengineering and Electronics of Russian Academy of Sciences, Mokhovaya 11-7, Moscow 125009, Russia
}%

\date{\today}% It is always \today, today,

\begin{abstract}
The recent \New{demonstrations} of viscous hydrodynamic electron flow in two-dimensional \New{electron} systems poses serious questions to the validity of existing transport theories, including the ballistic model, the collision-induced and collisionless hydrodynamics. While the theories of transport at hydrodynamic-to-ballistic crossover for free 2d electrons are well established, the same is not true for electrons in magnetic fields. In this work, we develop an analytically solvable model describing the transition from ballistic to hydrodynamic transport with changing the strength of electron-electron collisions in magnetic fields. Within this model, we find an expression for the high-frequency non-local conductivity tensor of 2d electrons. It is valid at arbitrary relation between frequency of external field $\omega$, the cyclotron frequency $\omega_c$, and the frequency of e-e collisions $\tau^{-1}_{ee}$. We use the obtained expression to study the transformation of 2d magnetoplasmon modes at hydrodynamic-to-ballistic crossover. In the true hydrodynamic regime, $\omega\tau_{ee} \ll 1$, the 2DES supports a single magnetoplasmon mode that is not split at cyclotron harmonics. In the true ballistic regime, $\omega\tau_{ee} \gg 1$, the plasmon dispersion develops splittings at cyclotron harmonics, forming the Bernstein modes. A formal long-wavelength expansion of kinetic equations (''collisionless hydrodynamics'') predicts the first splitting of plasmon dispersion at $\omega\approx 2\omega_c$. Still, such expansion fails to predict the zero and negative group velocity sections of true magnetoplasmon dispersion, for which the full kinetic model is required.
\end{abstract}

\maketitle

%\tableofcontents

\section{\label{sec:level1} Introduction}

In common semiconductor materials, scattering of electrons by phonons and impurities leads to the diffusive Ohmic transport. At the same time, in pure materials, the characteristic momentum relaxation length $l_{mr}$ for scattering by phonons and impurities can exceed the size of the device $W$. This  leads to  ballistic dc transport  
%at a high length of electron-electron collisions $l_{ee} \ll W$ 
with predominant scattering of electrons at the sample boundaries~\cite{ball_mayorov2011micrometer,ball_heremans1992}. A very interesting situation appears when momentum-conserving electron-electron collisions are intense, so that the respective free path $l_{ee} \ll l_{mr}$ and $l_{ee} \ll W$. In this case, electrons form a viscous fluid with its dynamics described by the Navier-Stokes equations.  This corresponds to the hydrodynamic dc transport, where current whirlpools and negative nonlocal resistance~\cite{hyd_gurzhi1968hydrodynamic,hyd_levitov2016electron,hyd_torre2015nonlocal, hyd_pellegrino2016electron,hyd_chandra2019hydrodynamic} can be observed. This hydrodynamic regime in electronic systems has
attracted significant interest and has been experimentally demonstrated in graphene~\cite{gr_bandurin2016negative, gr_kumar2017superballistic, gr_bandurin2018fluidity,gr_berdyugin2019measuring,gr_ku2020imaging}, (Al,Ga)As heterostructures~\cite{AlGaAs_braem2018scanning}, GaAs quantum wells~\cite{GaAs_gusev2018viscous, GaAs_gusev2021viscous} and Weyl semimetals~\cite{Ws_gooth2017experimental,Ws_gooth2018thermal,Ws_osterhoudt2021evidence}. 

The parameter space for diffusive, ballistic, and hydrodynamic regimes changes at finite frequency of external field $\omega$. A number of interesting effects in ac transport was predicted theoretically for ideal electron fluid~\cite{Hd_moessner2018pulsating,Hd_semenyakin2018alternating, Hd_cohen2018hall,Hd_alekseev2018magnetic} where e-e collisions are so intense that $\omega\tau_{ee} \ll 1$. \new{The ballistic regime in this case refers to a model of a collisionless electron plasma that is valid for describing processes that occur in times shorter than the free path of electrons ($\omega\tau_{ee} \gg 1$ and $\omega\tau_{mr} \gg 1$) or for processes whose characteristic spatial scales are smaller than the free path lengths ($q l_{ee} \gg 1$, $q l_{mr} \gg 1$, where $q$ - characteristic wave vector of wave processes)~\cite{aleksandrov1978principles}}.
%With increasing the frequency, the electrons' motion becomes effectively ballistic if no collisions appear during the field cycle, i.e. at $\omega\tau_{ee} \gg 1$ and $\omega\tau_{mr} \gg 1$.
%In the case of AC transport in pure materials, the ballistic and hydrodynamic regimes are determined by the ratio of the driving field frequency $\omega$ and the frequency of electron-electron collisions $\tau^{-1}$. At $\omega \tau_{ee} \ll 1$ and $\omega \tau_{ee} \gg 1$, we have ballistic  and hydrodynamic transport respectively. 
In many experimentally relevant situations, particularly, at THz or GHz frequencies, $\omega$ and $\tau_{ee}^{-1}$ are comparable in magnitude. This leads to the need to construct models of the transport intermediate between
hydrodynamic and ballistic regimes~\cite{IR_muller2009graphene,IR_muller2008quantum,IR_narozhny2015hydrodynamics,IR_foster2009slow,IR_svintsov2012hydrodynamic,IR_lucas2016hydrodynamic,IR_mayorov2011micrometer,HBcross,IR_chandra2019hydrodynamic,IR_holder2019ballistic,IR_gupta2021hydrodynamic,IR_tomogr,IR_heremans1992observation}.
%\begin{table}
%\caption{Values of parameters corresponding to the considered electronic transport regimes}
%\label{table}
%\begin{tabular}{| l | l |}
%\hline
%AC transport regime & Conditions  \\ \hline
%Ballistic & \omega \tau_{ee} \gg 1, \omega \tau_{mr} \gg 1 \\ \hline
%"True" hydrodinamic &  \omega \tau_{ee} \ll 1, \omega \tau_{mr} \gg 1 \\ \hline
%"Collisionless" hydrodinamic & \omega \tau_{ee} \gg 1, \omega \tau_{mr} \gg 1, q R_c \ll 1 \\
%\hline
%\end{tabular}
%\end{table}
It's important to mention that {\it ideal} electron fluid is formed at very strong e-e collisions, contrary to ideal electron gas where electrons do not interact at all. Indeed, the dc kinematic viscosity $\nu_{kin} = v_0^2 \tau_{ee}/4$ ($v_0$ is the Fermi velocity) tends to zero for very short free path time $\tau_{ee}$. \new{On the other hand, the ballistic transport can be equivalently termed as 'highly viscous'}. The true Navier-Stokes hydrodynamics is inapplicable in this regime. Yet it is possible to expand the transport equations in the long-wavelength limit, generating the nearly local 'collisionless hydrodynamics' or 'hydrodynamics of highly viscous fluid' \new{in which the equations of motion can be interpreted as the Navier-Stokes equations}. ~\cite{aleksandrov1978principles, Hd_alekseev2018magnetic}.

Such equations predict the existence of a novel collective mode, the transverce zero sound, in the absence of a magnetic field~\cite{Hd_lucas2018electronic, Hd_khoo2019shear}. In finite magnetic field $B$, the dynamic viscosity $\eta(\omega,B)$ was predicted to have the viscoelastic resonance~\cite{Hd_alekseev2018magnetic,alekseev2019,alekseev2019magnetosonic}. The latter should occur at double cyclotron frequency $\omega=2\omega_c$. Interaction between 2d plasmons and viscoelastic resonance was predicted to lead to emergence of novel collective mode, dubbed as "transverse zero magnetosound"~\cite{alekseev2019}. 

The predicted mode structure of highly non-ideal electron fluid contradicts to that of ballistic electrons, though both regimes should be equivalent and appear at $\omega\tau_{ee}\gg 1$. The ballistic theory states that plasmon dispersion interacts with multiple cyclotron resonances forming the so called Bernstein modes~\cite{BM_chiu1974plasma,BM_bernstein1958waves,BM_sitenko1957oscillations,BM_batke1985nonlocality,BM_gudmundsson1995bernstein,BM_batke1986plasmon,BM_holland2004quantized, BM_roldan2011theory,BM_volkov2014bernstein,bandurin2021}. The fundamental Bernstein mode, like conventional magnetoplasmon, starts from $\omega_c$ at $q=0$. As it approaches $2\omega_c$, it reaches a plateau, and then falls downward with negative group velocity. The group velocity on the plateau is zero, and density of states is singular. The plateau frequency is shifted from $2\omega_c$ by minigap. The next branch of the Bershtein modes starts from $2 \omega_c$ at $q=0$ and forms a plateau near the triple cyclotron frequency. Similar branches occur at each multiple of the cyclotron frequency.

In this \New{article}, we are aiming to resolve the abovementioned contradictions between 'ballistic' and 'highly-viscous hydrodynamic' models for 2DES in magnetic field. \new{To this end, we develop a generalized classical kinetic model that describes magnetotransport in ballistic and hydrodynamic transport regimes as limiting cases} (with $\varepsilon_F \gg T$). The model is based on \New{Boltzmann} kinetic equation with model \New{momentum and particle conserving} e-e collision integral. We calculate the nonlocal high-frequency conductivity with non-zero magnetic field, which is a necessary block for calculating many light-matter interaction characteristics~\cite{Sig_ciraci2012probing,Sig_gonccalves2021quantum,Sig_lundeberg2017tuning} (Section~\ref{Sec:conductivity}).

Using the expression for conductivity, we study the evolution of the dispersion of magnetoplasmons during the transition from the hydrodynamic regime to the ballistic regime (Section~\ref{Sec:MP_dispersion}). At a high frequency of electron-electron collisions corresponding to the hydrodynamic regime, an conventional magnetoplasmon with a single cyclotron gap is observed. As the collision frequency decreases, more and more pronounced dispersion splittings at multiple cyclotron frequencies appear, which in the ballistic regime take on the explicit form of Berstein modes. We inspect the approximation of "collisionless" hydrodynamics~(Section~\ref{Sec:collisionless_HD}) and find that is applicable only for wave vectors smaller than the anticrossing point $q^*$ of the two lowest Bernstein modes, coinciding with them in the region of applicability. This means that transverse zero
magnetosound is an artifact of long-wave expansion of transport equations beyond the region of applicability.

\begin{figure}[ht!]
\center{\includegraphics[width=1\linewidth]{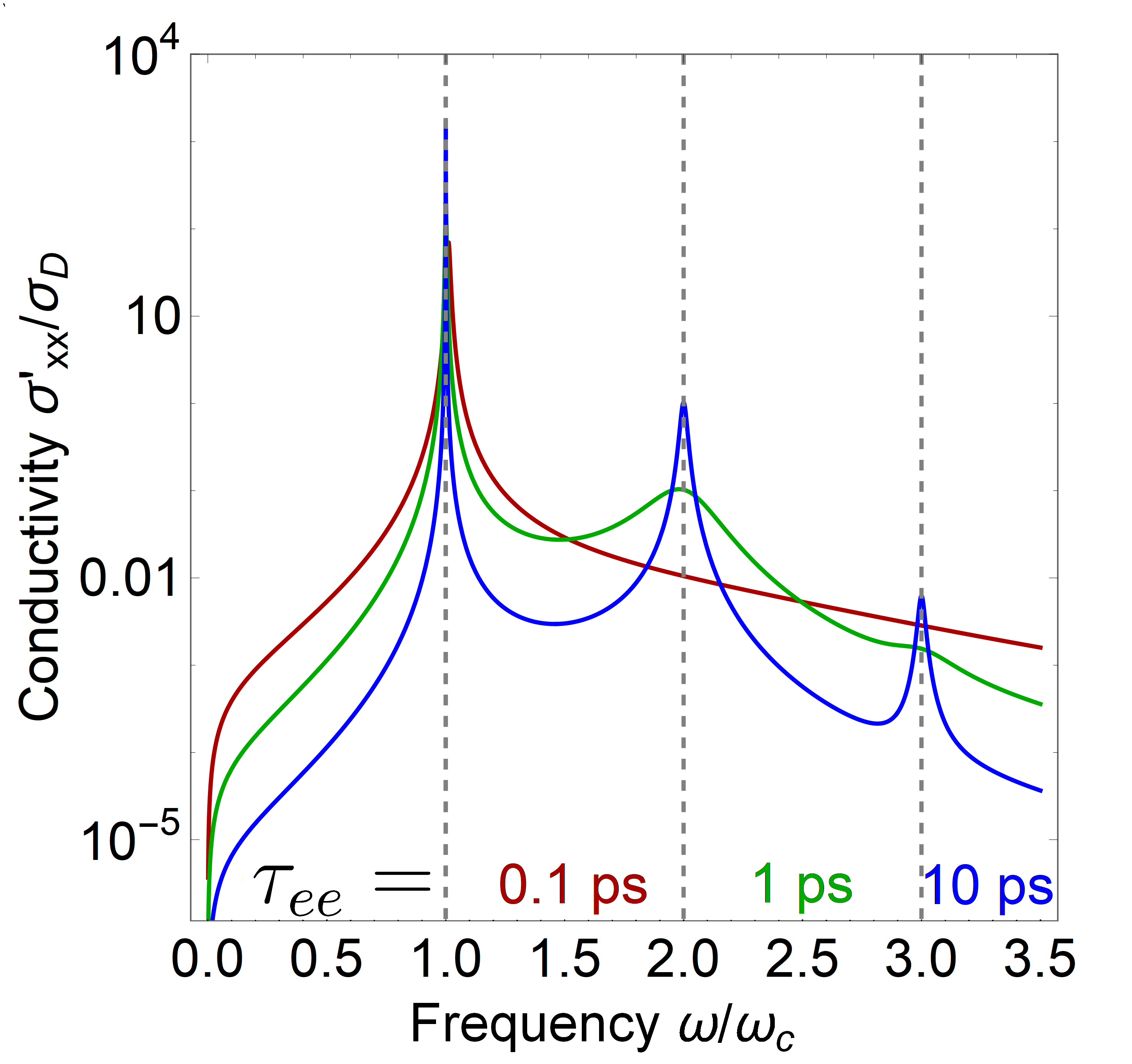} }
\caption{Real part of longitudinal conductivity $\sigma'_{xx}$ in the hydrodynamic regime with $\tau_{ee} = 0.1$ ps (red line), intermediate regime with $\tau_{ee} = 1$ ps (green line), and in the  ballistic regime with $\tau_{ee} = 10$ ps (blue line). The conductivity is normalized by collisionless Drude conductivity $\sigma_D = i n e^2/m \omega$. All curves are obtained at  carrier density $n_0 = 2 \times 10^{12}$ cm$^{-2}$,  $f_c = 1$ THz, and wave vector $q = 1$ $\mu$m$^{-1}$, $q R_c \approx 6.3$.}
\label{Conductivity}
\end{figure}  

\section{\label{Sec:conductivity} Non-local high frequency conductivity tensor}

We start with evaluation of the non-local conductivity tensor of 2DES $\hat\sigma({\bf q},\omega)$ in a classically strong perpendicular magnetic field $B \parallel \mathbf{e}_z$. To this end, we find the current response in a weak electric field $\delta {\bf E} = \delta \mathbf{E}_0 e^{i (\mathbf{q r} - \omega t)}$ with $\mathbf{r}, \mathbf{q} \perp \mathbf{e}_z$. It is found from the solution of linearized kinetic equation
\begin{equation}
     -(i \omega - i \mathbf{q v_p}) \delta f - e \delta \mathbf{E_0 v_p}  \dfrac{\partial f_0}{\partial \mathbf{\epsilon}} + \dfrac{1}{c} \mathbf{v_p} \times \mathbf{B}   \dfrac{\partial \delta f}{\partial \mathbf{p}}  = {\rm St}_{ee}\{\delta f\},
\label{KinEq}
\end{equation}
where distribution function $f = f_0 + \delta f e^{i (\mathbf{q r}-\omega t)}$, ${\bf v_p}$ is the electron velocity, $c$ the speed of light and ${\rm St}_{ee}\{\delta f\}$ the electron-electron collision integral. In a practically important limit $\varepsilon_F \gg T$ \new{in which the electron-phonon scattering can be neglected~\cite{narozhny2021hydrodynamic}}, only electrons with velocities close to the Fermi velocity $v_F$ participate in high-frequency kinetic processes, which makes this model applicable for describing both classical 2DES and graphene. Contrary to electron-impurity and electron-phonon collisions, e-e collisions conserve the net momentum of colliding particles. To account for this fact, we adopt the e-e collision integral in generalized relaxation-time approximation~\cite{HBcross,KE_abrikosov1959theory,KE_bhatnagar1954model,KE_gross1956model,KE_mermin1970lindhard,KE_lucas2018kinetic,KE_guo2017higher,KE_lucas2017stokes}. The collision integral pushes the distribution function $\delta f$ towards local equilibrium
\begin{equation}
\label{EE-collision-integral}
     {\rm St}_{ee}\{\delta f\} = \frac{\delta f- \delta f_{hd}}{\tau_{ee}}, \ \ \  \delta f_{hd} = -\frac{\partial f_0}{\partial \varepsilon}\left(\delta \mu + \mathbf{p} \delta \mathbf{v}\right),
\end{equation}
where $\delta \mu$ and $\delta \mathbf{v}$ are found from the conservation of particle number and momentum: 
\begin{gather}
    \label{ConsLaw}
    \sum_{\mathbf{p}} \left( \delta f - \delta f_{hd} \right) = 0,  \qquad
    \sum_{\mathbf{p}} \mathbf{p} \left( \delta f - \delta f_{hd} \right) = 0.
\end{gather}

The kinetic equation with e-e collision integral (\ref{EE-collision-integral}) and conservation laws (\ref{ConsLaw}) are sufficient to describe the behaviour of electron liquid at large wave vectors $q\sim \omega/v_0$,  and across the whole hydrodynamic-to-ballistic crossover. The scheme of solution is described in Appendix A. First, one passes to the polar coordinates in momentum space and expands the distribution function in Fourier series. Physically, the coefficients of these series are cyclotron harmonics. The kinetic equation in cyclotron harmonic representation is purely algebraic. Its symbolic solution at the first stage contains two unknown variables, $\delta\mu$ and $\delta{\bf v}$. On the second stage, these 'hydrodynamic quantities' are found from the conservation laws~(\ref{ConsLaw}). In a practically important limit $\varepsilon_F \gg T$, this leads us to a system of linear equations describing both ballistic and hydrodynamic behavior of 2DES:
\begin{widetext}
\begin{equation}
\left( \begin{array}{ccc}
1 - i \gamma Y^{(0)}_{00} & -2 i \gamma Y^{(1)}_{00} & -2 \gamma Y^{(0)}_{10} \\
- i \gamma Y^{(1)}_{00} & 1 - 2 i \gamma Y^{(2)}_{00} & -2 \gamma Y^{(1)}_{10} \\ 
\gamma Y^{(0)}_{10} & 2 \gamma Y^{(1)}_{10} & 1 - 2 i \gamma Y^{(0)}_{11} \end{array} \right)
\left( \begin{array}{ccc}
\frac{\delta n}{n_0} \\
\frac{\delta v_x}{v_0} \\
\frac{\delta v_y}{v_0} \end{array} \right) = 
-\left( \begin{array}{ccc}
0 \\
i Y^{(2)}_{00} E_x +  Y^{(1)}_{10} E_y \\ 
i Y^{(0)}_{11} E_y - Y^{(1)}_{10} E_x \end{array} \right) \dfrac{2 e}{\omega m},
\label{GenHD}
\end{equation}
\end{widetext}
where $\gamma = (\omega\tau_{ee})^{-1}$ is the dimensionless collision frequency, $v_0$ the Fermi velocity, $n_0$ equilibrium carrier density and $\mathbf{q} \parallel \mathbf{e}_x$. The dimensionless factors
\begin{equation}
Y_{ij}^{(k)} = \sum_{s=-\infty}^{\infty} \left( \dfrac{s}{q R_c}\right)^k \dfrac{ J^{(i)}_s[q R_c]J^{(j)}_s[q R_c]}{1 - s\omega_c / \omega + i \gamma},
\end{equation}
where $J^{(i)}_s[x]$ is the $i$-th order derivative of the Bessel function, $R_c = v_0/ \omega_c$ a cyclotron radius and $\omega_c$ a cyclotron frequency.

These equations can be interpreted as generalized hydrodynamics; the term 'generalized' implies applicability at arbitrarily large wave vectors. When expanded to terms $\propto q^2$ (with $qR_c \ll 1$), we obtain "collisionless" hydrodynamics ($ \omega \tau_{ee} \gg 1$) or "true" hydrodynamics ($ \omega \tau_{ee} \ll 1$):
\begin{eqnarray}
- i \omega \delta n + i (\mathbf{q}  \delta \mathbf{v}) n_0 = 0, 
\label{cont}
\end{eqnarray}
\begin{eqnarray}
-i \omega \delta \mathbf{v} = \omega_c [\delta \mathbf{v} \times \mathbf{e_z}] - i \mathbf{q} \dfrac{ v_0^2}{2} \delta n - \eta_{xx}(\omega) q^2 \delta \mathbf{v} - \nonumber\\ \qquad \qquad \qquad \qquad -\eta_{xy}(\omega) q^2 \delta [\mathbf{v} \times \mathbf{e_z}] + \dfrac{e \delta \mathbf{E}_0 }{m},
\label{Nav}
\end{eqnarray}
where the ac shear viscosity coefficients $\eta_{xx}(\omega)$ and $\eta_{xy}(\omega)$ depend on magnetic field and frequency as 
\begin{eqnarray}
   \eta_{xx}(\omega) = \eta_0 \dfrac{1-i \omega \tau_{ee}}{(1-i \omega \tau_{ee})^2 + 4 \omega_c^2}, \\
   \eta_{xy}(\omega) = \eta_0 \dfrac{2 \omega_c \tau_{ee}}{(1-i \omega \tau_{ee})^2 + 4 \omega_c^2},
\end{eqnarray}
where $\eta_0 = v_0^2 \tau_{ee}/4$ is the viscosity in the absence of magnetic field. Whereas "true" hydrodynamics, due to the small relaxation time, is also valid for large wave vectors, collisionless hydrodynamics has a limited range of applicability ($q v_0 \ll \omega_c$) which requires special care when applying this approximation for large wave vectors.

Finally, the components of the conductivity tensor $\hat{\sigma}(\omega, q)$ can be found from
\begin{equation}
    {\bf j} = e n_0 {\delta \bf v} = \hat\sigma ({\bf q},\omega) \delta{\bf E}_0,
\label{def}
\end{equation}
where the relation between ${\delta \bf v} $ and $\delta{\bf E}_0$ is found from \New{equations} (\ref{GenHD}) . 

It is instructive to track the changes in longitudinal conductivity $\sigma_{xx}({\bf q},\omega)$ while varying the strength of e-e collisions. Fig.\ref{Conductivity} illustrates this evolution and shows the frequency dependence of ${\rm Re}\sigma_{xx}$ at fixed $q= 1 \ \mu m^{-1}$ and $\omega_c/2\pi=1$ THz. In the ballistic regime ($\tau_{ee}=10$ ps, blue curve), the conductivity has resonances at multiple cyclotron frequencies which amplitude decreases with increasing the resonance order. As the frequency of e-e collisions $\tau_{ee}^{-1}$ becomes comparable to $\omega_c$, multiple resonances rapidly fade away. Only the main cyclotron resonance persists in the hydrodynamic regime, and its width shrinks again at very short values of $\tau_{ee}$, i.e. at low viscosity.

\section{\label{Sec:MP_dispersion}Magnetoplasmon dispersion}

\begin{figure*}[ht!]
\center{\includegraphics[width=1\linewidth]{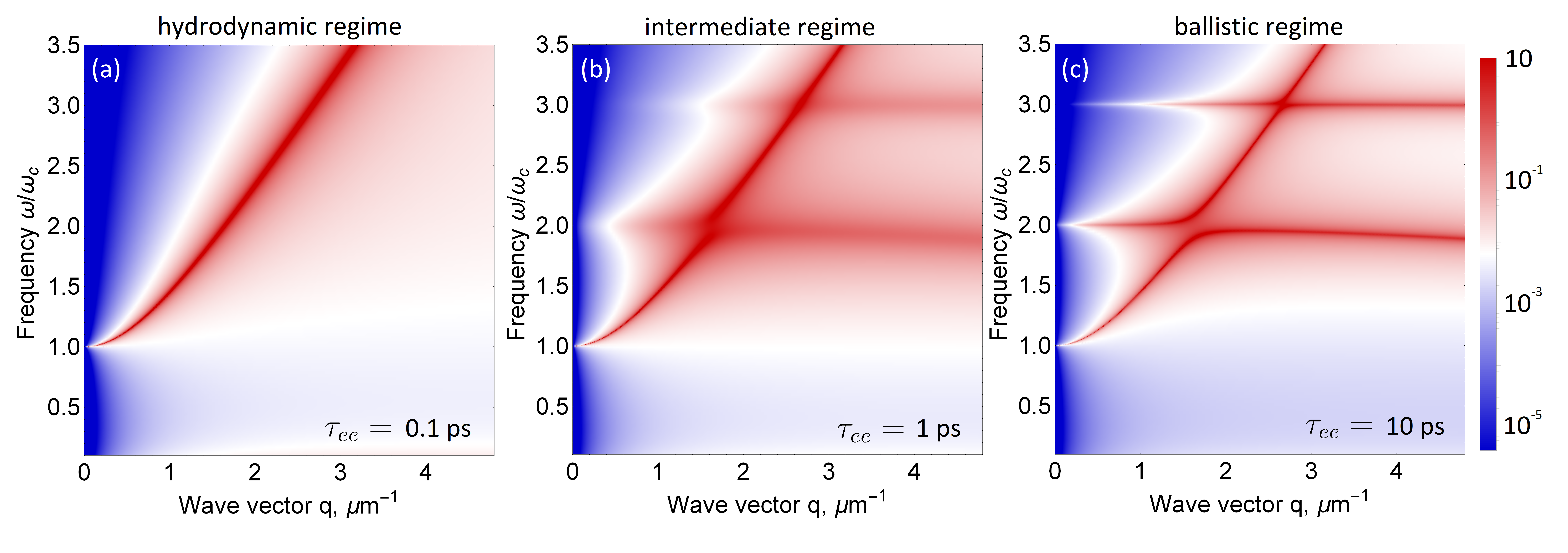} }
\caption{Magnetoplasmon  dispersion curve visualized through loss function ${\rm Im }\varepsilon^{-1}(q,\omega,\omega_c)$ in (a) the hydrodynamic transport regime in graphene, $\tau_{ee} = 0.1$ ps, (b) intermediate regime, $\tau_{ee} = 1$ ps, (c) in  ballistic regime, $\tau_{ee} = 10$ ps. All curves are obtained at  carrier density $n_0 = 2 \times 10^{12}$ cm$^{-2}$,  $f_c = 1$ THz with gate distance $d = 100$ nm, $\chi$ = 4}
\label{BM_and_hydrodynamics}
\end{figure*}  

We proceed to inspect the evolution of collective excitations in magnetized 2DES across the hydrodynamic-ballistic crossover. To address this problem, we consider the excitation of collective modes by a point dipole located above the 2DES in the quasi-static regime ($\omega/q \ll c$). We find that total electric potential in the 2DES plane can be represented as (Appendix B):
\begin{equation}
    \varphi({\bf q},\omega)  = \frac{\varphi_0({\bf q},\omega)}{\varepsilon({\bf q},\omega)},
\end{equation}
where $\varphi_0({\bf q},\omega)$ is the electric potential of the point source in the absence of 2DES and $\varepsilon({\bf q},\omega)$ the 2DES dielectric function. In the quasi-static regime, it is affected only by the longitudinal component of the conductivity tensor $\sigma_{xx}({\bf q},\omega)$ and is given by
\begin{equation}
    \varepsilon(\omega, q) = 1 + i \dfrac{2 \pi q}{\chi \omega} (1 - e^{-2 q d}) \sigma_{xx}({\bf q},\omega),
\label{diel}
\end{equation}
where $\chi$ is the effective dielectric constant, $d$ the distance to the gate electrode located below the 2DES and the longitudinal conductivity $\sigma_{xx}$ was obtained before in Sec.~\ref{Sec:conductivity}.

It is convenient to visualize the dispersion of plasmons using the loss function ${\rm Im}\varepsilon^{-1}$. The peaks in loss function correspond to a resonant enhancement of external electric field, i.e. to the collective electrostatic waves -- (magneto)plasmons. Using the obtained generalized equations of hydrodynamics, it is possible to trace the evolution of collective modes with an increase in the frequency of electron-electron collisions up to the transition to the hydrodynamic regime.

The emerging loss functions are displayed in Fig.~\ref{BM_and_hydrodynamics}. Our calculations show that the splittings of plasmon dispersion at multiple cyclotron frequencies are gradually blurred with increasing  e-e collision frequency $\tau_{ee}^{-1}$. In the hydrodynamic regime, the dispersion collapses into an ordinary magnetoplasmon
\begin{equation}
    \omega^2 = \omega_c^2 + s(q)^2 q^2,
\label{MP}
\end{equation}
where $s(q)= \sqrt{\dfrac{2 \pi n_0 e^2}{m \kappa q} (1-e^{-2 q d})}$.

%\New{This trend has some experimental evidence: the terahertz photoresponse of two-dimentional electron system in graphene located in perpendicular magnetic field shows a resonant burst at multiple cyclotron frequencies, which sharply exceeds the signal detected at the site of a conventional CR. Qualitatively, the effect can be described as follows: an incident electromagnetic wave diffracts on sharp, highly conductive contacts, creating a highly inhomogeneous near field, which excites the Bernstein modes in graphene. Each spatial harmonic of this inhomogeneous field is screened by graphene electrons, and in the vicinity of the magnetoplasmon dispersion, screening acquires amplifying properties. The amplification becomes especially strong in the vicinity of dispersion points with zero group velocity, which are located close to multiple cyclotron frequencies. This effect is observed only in very pure materials with high mobility, where the ballistic transport model is applicable~\cite{bandurin2021,dai2010}.}

\begin{figure}[ht!]
\center{\includegraphics[width=1\linewidth]{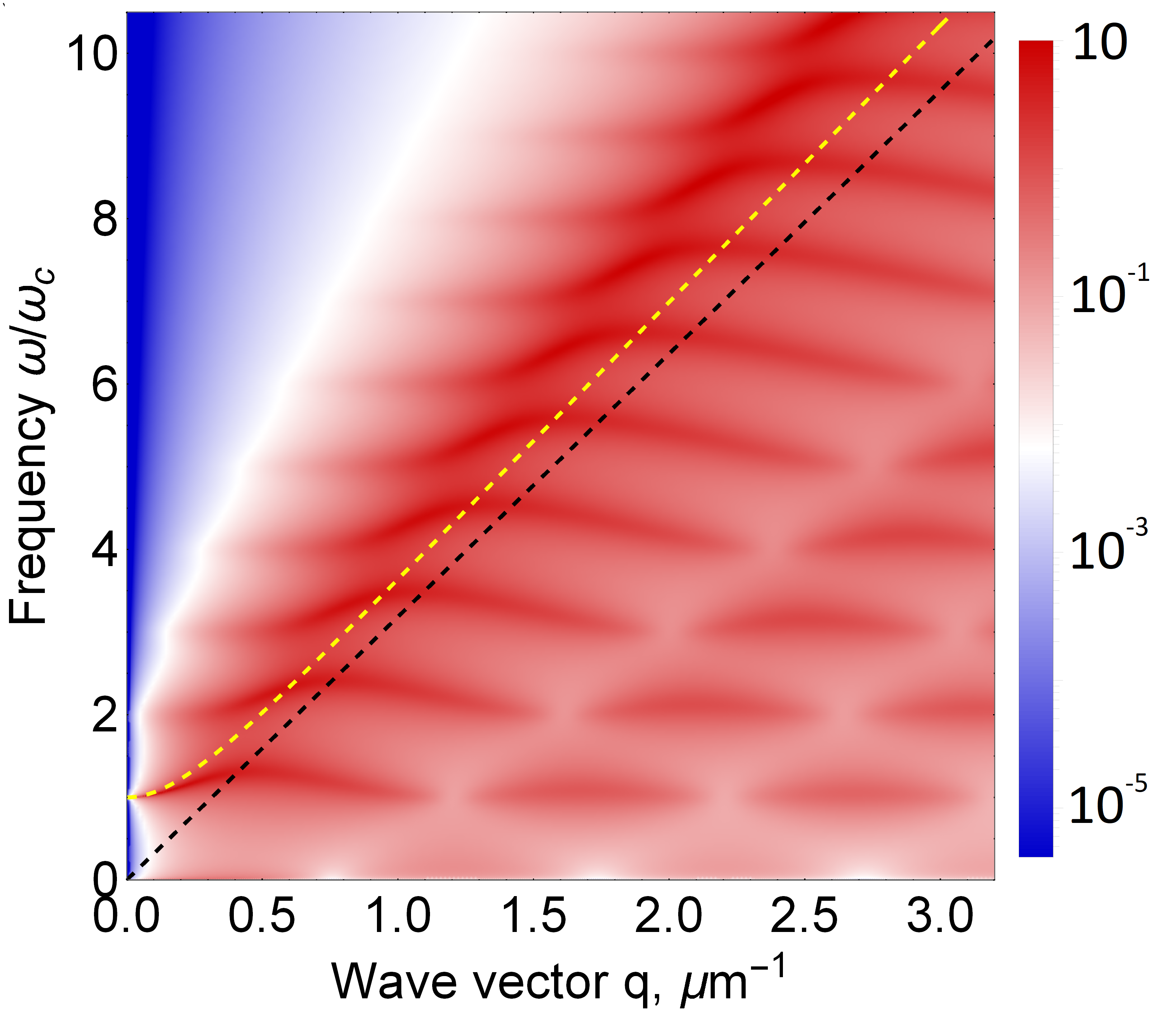} }
\caption{Magnetoplasmon  dispersion curve in graphene visualized through loss function ${\rm Im }\varepsilon^{-1}(q,\omega,\omega_c)$ with $\tau_{ee} = 20$ ps, carrier density $n_0 = 6.7 \times 10^{9}$ cm$^{-2}$,  $f_c = 0.05$ THz with gate distance $d = 50$ nm, $\chi$ = 4. Black dashed line shows the boundary of the domain of Landau damping, yellow dashed line shows the conventional magnetoplasmon dispersion curve.}
\label{BoundDestr}
\end{figure}  

Now it's important to track the transition from the obtained results to the zero magnetic field. In the absence of a magnetic field in the ballistic regime, the phase velocity of plasmons does not fall below the Fermi velocity in Dirac materials; thus, the dispersion of plasmons does not fall into the Landau damping region, and at low values of the carrier density $n_0$ and a low gate distance $d$, the dispersion asymptotically tends to the boundary $\omega = q v_0$~\cite{HBcross,lundeberg2017tuning}. However, in the presence of a weak magnetic field, this boundary is destroyed due to the splitting of the Bershtein modes (Fig. \ref{BoundDestr})

\begin{figure*}[ht!]
\center{\includegraphics[width=1\linewidth]{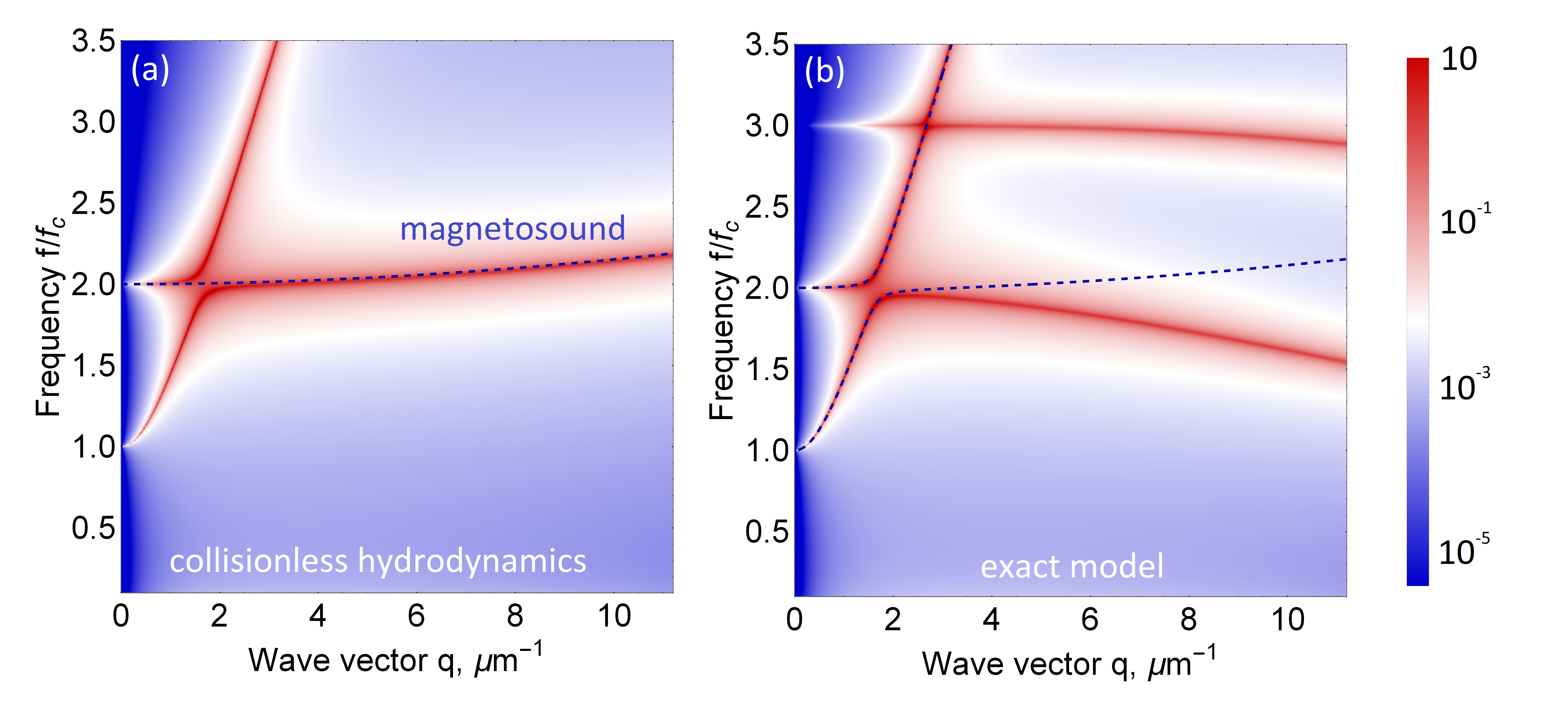} }
\caption{Magnetoplasmon  dispersion curve in graphene visualized through loss function ${\rm Im }\varepsilon^{-1}(q,\omega,\omega_c)$ in (a) the "collisionless" hydrodynamics model, dashed line shows the asymptotics of the lowest mode corresponding to the transverse magnetosound mode $\omega^2 = 4 \omega_c^2 + u_0^2 q^2 / 4$; (b) in the full theory, dashed lines correspond to modes obtained in the "collisionless" hydrodynamics approximation. All curves are obtained at  carrier density $n_0 = 2 \times 10^{12}$ cm$^{-2}$,  $f_c = 1$ THz with gate distance $d = 100$ nm, $\chi$ = 4 and $\tau_{ee} = 10$ ps}
\label{MS_BM}
\end{figure*}  

\section{\label{Sec:collisionless_HD}Collisionless hydrodynamic model}

 In this section, we study the range of applicability of collisionless hydrodynamics for the description of magnetoplasmon modes. Using the continuity equation (\ref{cont}) and the Navier-Stokes equations (\ref{Nav}), we find the conductivity of a two-dimensional system by definition (\ref{def}). We plot the loss function in the collisionless hydrodynamic model (\ref{cont}-\ref{Nav}) and in the ballistic transport model (\ref{GenHD}) (Fig.\ref{MS_BM}). In the first one (Fig.\ref{MS_BM}a) the zeros of the dielectric function (\ref{diel}) in this case give two magnetoplasmon modes. The first of them goes from the cyclotron frequency $\omega_c$, at small wave vectors it coincides with the conventional magnetoplasmon (\ref{MP}), and at large wave vectors (after the anti-crossing point $q^{*}$) it coincides with the transverse magnetosound
\begin{equation}
    \omega^2 = 4 \omega_c^2 + \dfrac{v_0^2 q^2}{4},
\label{MS}
\end{equation}
which is related to perturbations of shear stress of a charged Fermi liquid in a magnetic field.~\cite{alekseev2019} The second mode starts from the doubled cyclotron frequency and after the anti-crossing point tends to conventional magnetoplasmon. It is possible to approximately estimate the anti-crossing point $(\omega^*, q^*)$ as the point at which the value of doubled cyclotron frequency $2 \omega_c$ is reached by the dispersion of conventional magnetoplasmon (\ref{MP}): 
\begin{equation}
    \omega^* \approx 2 \omega_c, \ \ \ \ \ \ \ q^* \approx \dfrac{\sqrt{3} \omega_c}{s}.
\label{antiC}
\end{equation}
In this case, up to the anti-crossing point $q^*$, there is a good agreement with the two lowest Bernstein modes observed in the full theory (Fig.\ref{MS_BM}b) (but modes starting at multiple cyclotron frequencies are not observed). However, for large wave vectors, serious discrepancies are observed. In the collisionless hydrodynamics model, the lowest mode exceeds twice the cyclotron frequency and coincides with the magnetosonic mode, while in the full theory, \New{the fundamental} Bernstein mode never exceeds twice the cyclotron frequency and forms a plateau with a singular density of states.

\section{Discussion and conclusion}

In the limit $\omega \tau_{ee} \gg 1$, the constructed full kinetic theory describes Bernstein modes. Magnetosound waves predicted in the framework of collisionless hydrodynamics described by the equations (\ref{Nav}) in Ref.~\cite{alekseev2019} coincide with the full theory only in a narrow $q$-range. Since the lowest mode in the collisionless hydrodynamics model tends to magnetosound mode (\ref{MS}) only after the anti-crossing point (\ref{antiC}), where this model is inapplicable, it can be argued that transverse magnetosound modes are an artifact of spectrum cutoff in generalized hydrodynamic equations (\ref{GenHD}) and are not observed in the full theory. 

While the physics behind these differences is clear, it's interesting to see and compare the emerging absorption spectra. The dielectric function $\varepsilon(q,\omega,\omega_c)$ corresponds to the screening by electrons of the two-dimensional system of the external field (Appendix B). The loss function Im$[\varepsilon^{-1}]$ is responsible for the magnetoplasmon-assisted absorption. It peaks at the magnetoplasmon modes, and therefore should enhance the near-field magnetoabsorption of an inhomogeneous field incident on the 2DES.~\cite{bandurin2021} 

Both transverse magnetosound and Bernstein modes are predicted in the loss function (Fig.\ref{MS_BM}). In both cases, for pure materials, an asymmetric (with respect to the magnetic field) resonance enhancement of magnetoabsorption in the vicinity of the doubled cyclotron frequency is predicted. However, in the case of magnetosound, the sharp side of the resonance is located in the direction of high magnetic fields, and in the case of Bershtein modes, on the contrary, in the direction of low magnetic fields, which makes it possible to distinguish one from the other in measurements. 

Measurements of the photoresistence demonstrate similar asymmetric resonances at double, triple, and even quadruple the cyclotron frequency in pure graphene in a magnetic field~\cite{bandurin2021} that matches the Bernstein modes seen in the full theory and are not predicted by the model of collisionless hydrodynamics. In the full theory the sharp side of the resonance is located in the direction of low magnetic fields.  In Ref.~\cite{alekseev2019} it was shown for a thin strip of a 2DES in the case of collisionless hydrodynamics that the sharp side of the resonance is located in the direction of high magnetic fields. However, the shape of the resonance strongly depends on the width of the strip and becomes almost symmetric in the case of an infinite 2DES. It can be assumed that in this case the shape of the resonance should strongly depend on the shape of the sample. However, measurements of various samples show an asymmetry consistent with the full theory~\cite{dai2010,bandurin2021}, which is another argument that these resonances are caused by Bernstein modes.

%Measurements of the magnetoresistivity in ultrapure GaAs quantum wells based demonstrate an asymmetric resonance at twice the cyclotron frequency, with the asymmetry corresponding to the Bernstein modes~\cite{dai2010}.

In summary, a model of high-frequency dynamics of a two-dimensional electronic system has been constructed, which allows one to study the transition from the ballistic regime to the hydrodynamic. It is shown that the conventional magnetoplasmon in the hydrodynamic regime, with a decrease in the frequency of electron-electron collisions, acquires characteristic splittings at multiple cyclotron frequencies and becomes Bernstein modes in the ballistic limit. According to calculations, the model of ''collisionless'' hydrodynamics, which predicts transverse magnetosound, is applicable to the description of magnetoplasmon modes only for wave vectors not exceeding the anti-crossing point (\ref{antiC}) of the two lowest Berstein modes, and magnetosound mode are beyond the range of applicability of this model.

\begin{acknowledgments}
The work was supported by grant \# 21-12-00287 of the Russian Science Foundation \\
\end{acknowledgments}

\subsection*{Appendix A: Solution of the kinetic equation}
To solve the kinetic equation (\ref{KinEq}), we pass to polar coordinates (with $\mathbf{q} \parallel \mathbf{e}_x$) and represent the equation as:
\begin{widetext}
	\begin{gather}
	\label{KE_phi}
	 -i(\alpha + i \gamma_c) \delta f + i \beta \cos{\varphi} \delta f  + \dfrac{\partial \delta f}{\partial \varphi} =  \dfrac{\partial f_0}{\partial p} \left(  \dfrac{e E_x}{\omega_c}  \cos{\varphi} + \dfrac{e E_y}{\omega_c}  \sin{\varphi} + \dfrac{\gamma_c \delta \mu}{v_0} + \dfrac{\gamma_c p_F \delta v_x}{ v_0} \cos{\varphi} + \dfrac{\gamma_c p_F \delta v_y}{ v_0} \sin{\varphi}    \right),
	\end{gather}
\end{widetext}
where $\alpha = \omega / \omega_c$, $\beta = q R_c$, $\gamma_c = (\omega_c \tau_{ee})^{-1}$, and $\omega_c$ is the cyclotron frequency. Then, we seek for the function $\delta f$ in the form $\delta f = g e^{-i \beta \sin{\varphi}}$, and expand $g$ as a series of angular harmonics $\varphi$: 

\begin{equation}
 g = \sum_{s= -\infty}^{+\infty} g_s e^{i s \varphi}.   
\end{equation}
The solution for $g_s$ reads as:
\begin{eqnarray}
\label{gs}
	 g_s = i J_s(\beta) \dfrac{\partial f_0}{\partial p} \dfrac{ \left( \frac{e E_x}{\omega_c}  \frac{s}{\beta}  + \frac{\gamma_c \delta \mu}{v_0} + \frac{\gamma_c p_F \delta v_x}{v_0} \frac{s}{\beta} \right)}{\alpha + i \gamma_c - s}  + \nonumber\\
	 + J_s'(\beta) \dfrac{\left( \frac{e E_y}{\omega_c}+ \frac{\gamma_c p_F \delta v_y}{ v_0} \frac{s}{\beta}  \right)}{\alpha + i \gamma_c - s},    
\end{eqnarray}
where $J_s(\beta) = \dfrac{1}{2 \pi} \int_0^{2 \pi} e^{i(\beta \sin{\varphi} - s \varphi)} d \varphi$ is the $s$-th order Bessel function and $J'_s(\beta)$ its derivative. Using the equations of conservation of particles and momentum (\ref{ConsLaw}), we obtain the equations of high-frequency dynamics (\ref{GenHD})

\subsection*{Appendix B: Magnetoabsorption calculation}
\begin{figure}[ht!]
\center{\includegraphics[width=1\linewidth]{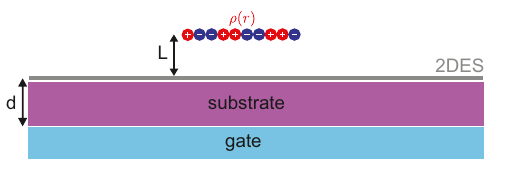} }
\caption{Schematic representation of a two-dimensional system, above which an inhomogeneous charge density is located near}
\label{Rho}
\end{figure}

In this section, we first consider the absorption of an inhomogeneous wave by an infinite two-dimensional electron system, then consider the case when an inhomogeneous field is created by planar charge density $\rho(\mathbf{r}) \sim e^{-i \omega t}$ located a short distance $L$ from the plane of a 2DES ($ qL \ll 1$) (Fig.\ref{Rho}). The absorption
given by the Joule's law: 
	\begin{gather}
	\label{Q}
	    P = 2 \int \Re{[\mathbf{E}(\mathbf{q},\omega) \mathbf{j}^*]}d \mathbf{q},
	\end{gather}
where $\mathbf{E}(\mathbf{q},\omega)$ is the total field in the
plane of the two-dimensional system.
Further, we take into account the Ohm's law  $\mathbf{j} = \hat{\sigma} \mathbf{E}$ with conductivity tensor \begin{equation}
\hat{\sigma} =
	  \begin{pmatrix}
                    \sigma_{xx} & -\sigma_{xy} \\
                    \sigma_{xy} & \sigma_{xx}
    \end{pmatrix}.  
\end{equation}
and obtain an expression for the absorbed power in the form
    \begin{gather}
        \label{Absorp}
        P = 2 \int \dfrac{d \mathbf{q}}{(2 \pi)^2} \left(\sigma'_{xx} |\mathbf{E}(\mathbf{q}, \omega)|^2 + \sigma''_{xy} \Im{[E^*_{ x} (\mathbf{q}, \omega)E_{y}({\mathbf{q}, \omega})]}\right).
    \end{gather}

In the plane of a two-dimensional system, the field is screened by electrons of 2DES. \New{In the electrostatic approximation ($ q c \gg \omega$), suitable for describing a sharp increase in absorption by a strongly inhomogeneous Bernstein mode field}, the total potential in the plane of the system is given by the expression

\begin{equation}
    \mathbf{\varphi}|_{z=0} =  \dfrac{2 \pi \rho(\mathbf{q})e^{-qL}}{\chi q \varepsilon({\bf q},\omega)},
\end{equation}
where $\varepsilon({\bf q},\omega)$ is the dielectric function and $\chi$ substrate dielectric constant.

%The incident field ${\bf E}_{inc}$ is screened by the electrons of the two-dimensional system, and the total field ${\bf E}$ in the plane of the two-dimensional system is determined by the relation $\mathbf{E}|_{z=0} = \mathbf{E}_{inc}|_{z=0}/\varepsilon(q,\omega,\omega_c)$, where $\varepsilon(q,\omega,\omega_c)$ is the dielectric function. A consistent way of obtaining $\varepsilon(\mathbf{q},\omega, \omega_c)$ is to consider the scattering of incident evancescent wave ${\bf E}_{inc} \propto e^{i q_x x + i q_y y} e^{-\sqrt{q^2 - k^2}z}$ by 2DES. In the quasi-static limit ($q \omega / c >> 1$, where $c$ is the speed of light) we replace the electric field in the plane of two-dimensional system $\mathbf{E}_d|_{z=0}$ by an equivalent distribution of surface charges $\rho(\mathbf{q})$ in the 2DES plane using the Gauss law:
%\begin{equation}
%    \mathbf{E}_{inc}|_{z=0} = - 2 \pi \rho(\mathbf{q})e^{-q %L} \dfrac{\mathbf{q}}{\chi q}.
%\end{equation}

To find the dielectric function $\varepsilon({\bf q},\omega)$ of a planar symmetric non-gated two-dimensional system  with a known conductivity tensor $\hat{\sigma}$, we solved the field equation (with $q L \ll 1$)
\begin{equation}
    \left( \dfrac{\partial^2 }{\partial z^2} - q^2 \right) \varphi(\mathbf{q},z) = \dfrac{4 \pi}{\chi} (\rho_{ind}(\mathbf{q}) + \rho(\mathbf{q})) \delta(z),
\end{equation}
where $\omega \rho_{ind}(\mathbf{q}) = \mathbf{q} \mathbf{j}$ and $\mathbf{j} = \hat{\sigma} \mathbf{E}$. Solving the obtained equations, we obtain an expression for the dielectric function
\begin{equation}
    \varepsilon({\bf q},\omega) = 1 + i \dfrac{2 \pi q}{\chi \omega} \sigma_{xx}(\omega, q).
\label{diel0}
\end{equation}
Taking into account the screening by a gate located at a distance $d$, we obtain the expression (\ref{diel})

In this case, the expression for absorption takes the form
\begin{equation}
    P_{\rm near} = 2\int{d{\bf q} \frac{\omega}{2 \pi \chi^2 q}|\rho(\mathbf{q})|^2 \Im \frac{1}{\varepsilon({\bf q},\omega)}}.
\end{equation}
The specific type of charge density $\rho(\mathbf{q})$ depends on the problem. Absorption caused by an inhomogeneous field of charges resulting from the diffraction of an incident plane wave on a thin contact was considered in detail in Ref.~\cite{bandurin2021}

%where ${\bf E}_{{\bf q}\omega} = F_{{\bf q}\omega} {\bf E}_{inc}$ is the amplitude of ${\bf q}$-th spatial harmonic of electric near field, which appeared, for example, as a result of the diffraction of the incident wave at the contact}

%\bibliography{my}% Produces the bibliography via BibTeX.


\end{document}